\documentclass[%
twocolumn,
superscriptaddress,
showpacs,
 amsmath,amssymb,
 aps,prb,
]{revtex4-1}
 
\usepackage{graphicx}
\usepackage{dcolumn}
\usepackage[section]{placeins}
\usepackage{bm}
\usepackage{color}
\usepackage{multirow}
\usepackage{ulem}
\usepackage{hyperref}
\usepackage{quantum}

\definecolor{green}{rgb}{0,0.7,0.3}

\definecolor{orange}{cmyk}{0.2,0.5,1.0,0.0}

\begin{document}
\widetext
\title{Stripe and superconducting order competing in the Hubbard model on a square lattice studied by a combined variational Monte Carlo and tensor network method}
YITP-18-92
\author{Andrew S. Darmawan}
\affiliation{Department of Applied Physics, The University of Tokyo, 7-3-1 Hongo, Bunkyo-ku, Tokyo 113-8656, Japan}
\affiliation{Yukawa Institute for Theoretical Physics (YITP), Kyoto University, Kitashirakawa Oiwakecho, Sakyo-ku, Kyoto 606-8502, Japan}
\author{Yusuke Nomura}
\affiliation{Department of Applied Physics, The University of Tokyo, 7-3-1 Hongo, Bunkyo-ku, Tokyo 113-8656, Japan}
\author{Youhei Yamaji}
\affiliation{Department of Applied Physics, The University of Tokyo, 7-3-1 Hongo, Bunkyo-ku, Tokyo 113-8656, Japan}
\affiliation{JST, PRESTO, 7-3-1 Hongo, Bunkyo-ku, Tokyo 113-8656, Japan}
\author{Masatoshi Imada}
\affiliation{Department of Applied Physics, The University of Tokyo, 7-3-1 Hongo, Bunkyo-ku, Tokyo 113-8656, Japan}

\date{\today}

\begin{abstract}
The long-studied Hubbard model is one of  the simplest models of copper-oxide superconductors.  However, the connection between the model and the experimental phase diagram is still under debate, in particular regarding the existence and extent of the $d$-wave superconducting phase. 
    Recent rapid progress in improving the accuracy of numerical solvers has opened a way to answer this question reliably. Here, we study the hole-doping concentration ($\delta$) dependence of the  Hubbard model in the ground states on a square lattice at strong coupling $U/t=10$, for the on-site interaction $U$ and the transfer $t$, using a variational Monte Carlo method. The method, which combines tensor network and Lanczos methods on top of Pfaffian wave functions, reveals a
rich phase diagram, in which many orders compete severely and degenerate within the energy range of 0.01$t$. 
We have identified
distinct phases including 
a uniform $d$-wave superconducting phase for $0.17\lesssim \delta \lesssim0.22$
and a stripe charge/spin ordered phase for $\delta\lesssim0.17$ with the stripe period depending on $\delta$, together with presumable spatially coexisting antiferromagnetic and stripe order for $\delta\lesssim0.07$ and coexisting stripe and $d$-wave superconductivity for $0.07\lesssim\delta\lesssim0.17$.
The present, improved method revealed a wider region of a charge uniform superconducting phase than the previous studies and shows a qualitative similarity to the phase diagram of the cuprate superconductors.
The superconducting order parameter is largest at doping of around $\delta=0.17$ in the ground state, which undergoes phase transitions from an inhomogeneous to a uniform state.
\end{abstract}

\maketitle

\section{Introduction.}
The mechanism of high-temperature superconductivity in doped Mott insulators remains a challenging open issue~\cite{anderson_resonating_1987,PhysRevB.37.3759,gros1988superconductivity,PhysRevLett.64.475,imada_metal-insulator_1998,lee_doping_2006}. In such systems, superconductivity severely competes with charge inhomogeneities~\cite{tranquada_evidence_1995,hoffman_four_2002,howald_coexistence_2003,neto_ubiquitous_2014,keimer_quantum_2015,comin_resonant_2016}, and resolving the different orders requires careful and accurate estimates. While a strong effective attractive interaction could be responsible for both Cooper pair formation and charge inhomogeneity (or the charge susceptibility enhancement)~\cite{PhysRevLett.64.475,PhysRevLett.67.259,nobuo_furukawa_two-dimensional_1992,emery_frustrated_1993,PhysRevLett.74.3652,PhysRevB.62.12700,watanabe_precise_2004,PhysRevB.78.165101,PhysRevLett.104.116402,PhysRevB.84.241110,misawa_origin_2014}, a true microscopic origin of these phenomena and their relationship is yet to be understood.

For this issue, an analysis of the single-band Hubbard model on a square lattice offers one of the simplest starting points. 
Many numerical studies~\cite{PhysRevLett.95.237001,PhysRevB.74.054513,PhysRevB.76.224509,PhysRevB.81.201101,PhysRevLett.108.216401,PhysRevB.86.241106,PhysRevB.88.245110,zheng_stripe_2017,ido_competition_2018}
 have shown that the Hubbard model is a unique playground which exibits
a large number of strongly competing orders observed in the cuprates, including $d$-wave superconducting and charge/spin stripe orders. 
Despite its simple form with only nearest-neighbor transfer ($t$) and on-site Hubbard interaction ($U$) terms, 
the ground-state phase diagram of the Hubbard model is still under active debate. 

Recent advancement of sophisticated numerical techniques and growing computational power have potentially brought the phase diagram of the Hubbard model within reach. 
The auxiliary-field quantum Monte Carlo method (AFQMC) \cite{sorella_numerical_1988, imada_numerical_1989} has enabled clarification of the instability to the charge inhomogeneity signaled by the diverging charge susceptibility near $\delta=0$ at a moderate coupling $U/t=4$ \cite{nobuo_furukawa_two-dimensional_1992,furukawa_charge_1993}. At stronger coupling ($U/t>6$), it has turned out that the charge instability with the critical divergence at $\delta=0$ is more expanded and the phase separation region occupies the low-doping region \cite{misawa_origin_2014}.
By allowing stripe-type charge inhomogeneity with the period of several lattice constants, a part of phase separation region at the strong coupling is replaced by the stripe ground state: A recent paper~\cite{zheng_stripe_2017} employed various numerical methods such as 
density matrix embedding theory (DMET)~\cite{PhysRevLett.109.186404},  
constrained path AFQMC~\cite{PhysRevLett.74.3652},
infinite projected entangled-pair states (iPEPS)~\cite{jordan_classical_2008},
and 
density matrix renormalization group (DMRG)~\cite{PhysRevLett.69.2863,PhysRevB.48.10345}
to study the ground state at 1/8 doping. 
This study mainly considered $U/t=8$, but less comprehensive results were also presented for $U/t=6$ and $U/t=12$.
All the methods support the stripe ground state with a near degeneracy of stripes with periods from 5 to 8. 

Systematic doping dependence has been studied by the many-variable variational Monte Carlo (mVMC) method~\cite{tahara_variational_2008-2,tahara_variational_2008} (see also  Refs. \onlinecite{sorella_generalized_2001,casula_geminal_2003,casula_correlated_2004,bajdich_pfaffian_2006} and a review \cite{gros_physics_1989})
and by various embedding methods~\cite{metzner_correlated_1989,muller-hartmann_hubbard_1989,brandt_thermodynamics_1989,georges_dynamical_1996,maier_quantum_2005,kotliar_electronic_2006, PhysRevLett.109.186404}. 
Among embedding methods, DMET was used to investigate the doping dependence for $U/t\le8$~\cite{zheng_ground-state_2016}. However, due to the restricted size of the embedded clusters used in that study, the roles of different period stripes could not be compared. Another embedding method, cellular dynamical mean-field theory, was also recently applied to the $U/t=6$ model\cite{vanhala_dynamical_2018}. While cluster sizes were allowed to vary to allow consideration of long stripe periods, certain discrepancies in this work compared to other works, for instance in the energetically preferred stripe periods \cite{zheng_stripe_2017}, suggests that conclusive results had not yet been reached.

The mVMC results for doping dependence suggest a charge/spin stripe ground state with the period increasing with decreasing $\delta$~\cite{ido_competition_2018}. 
However, these ground states are severely competing with superconducting excited states. 
Because the energy difference is tiny ($\sim 0.01 t$ per site, which corresponds to roughly $\sim  50$ K in the scale of the cuprates), improving the accuracy of wave functions might lead to different conclusions.

In this work, we use a method developed in Ref.~\onlinecite{zhao_variational_2017-1}, which combines the mVMC method with a tensor network (TN) method, to study the doping concentration dependence of the ground state in the strongly coupled $U/t=10$ Hubbard model, taking into account both homogeneous and inhomogeneous orders. 
The wave function is further improved with first-step Lanczos, and ground state energies are finally obtained by extrapolation to zero energy variance. 
In this way, we obtain energy estimates that are comparable to state-of-the-art calculations at $\delta=1/8$~\cite{zheng_stripe_2017}, and substantially better than the previous mVMC study~\cite{ido_competition_2018}.
Systems of various sizes are considered, and finite-size effects are carefully examined to estimate the thermodynamic limit. 

One of the most important findings of the present paper is that there is a stable uniform nonmagnetic ground state showing superconductivity in the doping region of $0.17\lesssim \delta\lesssim 0.22$, 
which has not been seen in the previous study~\cite{ido_competition_2018}. For the doping region with concentration between $0.07\lesssim \delta \lesssim 0.17$ 
we find stripe-ordered ground states. The superconductivity of stripe states in this region appears less robust than for the uniform region.
However, a possible region of phase separation for $0.12\lesssim \delta\lesssim 0.19$ suggests that a uniform state with higher doping
could largely account for the superconductivity in the doping range $0.12\lesssim \delta\lesssim 0.19$, where a mixture of stripe and
superconducting ordered domains could be stabilized.

\section{Definitions and method}
\label{s:def}
The target of the present study is the single-band Hubbard model on a square lattice, given by
\begin{equation}
    {\mathcal H}=-t\sum_{\langle i,j \rangle, \sigma} c_{i\sigma}^\dag c_{j \sigma} + U\sum_{i=1}^{N_s} n_{i{\uparrow}}n_{i{\downarrow}}\,,
\end{equation}
where the first sum is taken over all nearest-neighbor pairs, $c_{i\sigma}$ is the annihilation operator for an electron at site $i$ with spin $\sigma$, $n_{i\sigma}=c_{i\sigma}^\dag c_{i\sigma}$ is the corresponding number operator and $N_s=L_xL_y$ is the number of lattice sites. This study will primarily focus on the strong interaction regime with $U/t=10$ (except for benchmark calculations). 

At large $U/t$, the low energy space of the model is expected to host many different types of order which must be accurately described by the numerical method. For this purpose, we employ a tensor-network method on top of variational Monte Carlo method. 
The wave function can be written as~\cite{zhao_variational_2017-1} 
\begin{equation}
    \ket{\psi}=\sum_x \mathcal{P}(x) \mathcal{M}(x) \psi_{\rm pair}(x)\ket{x}\,,
    \label{e:wavefunction}
\end{equation}
where $\{\ket{x}\}$ is a basis of real-space configurations, $\psi_{\rm pair}(x)$ is a Pfaffian, $\mathcal{P}(x)$ are correlation factors, and $\mathcal{M}(x)$ is a fat tree tensor network. Full details of the definitions and roles of each factor are included in 
Appendix \ref{s:method}.
Further improvement to the wave function is obtained using first-step Lanczos \cite{heeb_systematic_1993} where the optimized variational wave function $\ket{\psi}$ is replaced with $\ket{\psi'}=(1+\alpha {\mathcal H})\ket{\psi}$ with a single variational parameter $\alpha$.

The ground state energy is arrived at by extrapolation with respect to the energy variance $\Delta_{\rm var}:=(\langle {\mathcal H}^2 \rangle - \langle {\mathcal H} \rangle^2)/\langle {\mathcal H} \rangle^2$, which is zero for the true ground state. Both the energy and energy variance are calculated for the optimized variational wave function with and without the tensor network factor, and with and without first-step Lanczos. 
Since linear relationship between variance and energy is shown if the obtained variational wave function is a good approximation containing substantial portion of the true ground state ~\cite{PhysRevB.48.12037,sorella_generalized_2001,kashima_path-integral_2001-1,PhysRevC.65.064319}, we regard the ground-state energy as the $y$-intercept for this linear fit of the above described four data points. See Ref.~\onlinecite{kashima_path-integral_2001-1} for the basis of the linearity.
In almost all of our calculations 
(as in Fig. \ref{f:varext} in Appendix \ref{s:method}), 
linear relationship between the energy and variance can been seen, resulting in a small fitting error. 
The only exception is seen close to the critical point in the uniform state.
 We remark that any deviation from this linear behavior that could occur at low energies is not taken into account in this method, and thus the linear approximation is a potential source of error not included in fitting error bars.
However, when the variance is small enough, we can expect that the error from the linear approximation is small. 

For most of our calculations we have used twisted boundary conditions which are periodic in the $y$-direction and anti-periodic in the $x$-direction.
This choice appears to improve the optimization and resulted in small finite-size effects. Inhomogeneous 
charge and spin
stripe orders could be optimized using a unit cell size commensurate with the period of the stripe. For stripes of odd charge period $\lambda$, a unit cell of size $\lambda\times 2$ was used, while for even period stripes a $2\lambda\times 2$ unit cell was used. 
This is because the length of spin period is twice larger than (equal to) that of charge when the charge period is even (odd)~\cite{white_density_1998,ido_competition_2018}. Examples of stripes with odd and even $\lambda$ are shown in Fig. \ref{f:phase_diagram}(a).
 Charge density variation is always chosen to be in the $x$ direction. For charge uniform states, a $2\times 2$ unit cell was used, which can accommodate antiferromagnetic and $d$-wave superconducting order. 

The system size was varied to examine finite-size effects and extrapolation of physical quantities. The largest systems used in our calculations were $24\times24$, $36\times16$ and $72\times 8$ which have $N_s=576$ sites, however certain quantities converged rapidly to thermodynamic limit (TL) values, and so only smaller sizes were required. Details of finite-size effects are included in 
Appendix \ref{s:finite}.

We have tested the accuracy of the method against various relevant benchmarks. Results of these benchmarks are provided in 
Appendix \ref{s:benchmarks}. 
At half-filling with $U/t=8$, a discrepancy of about $0.005t$ is observed in the energy per site, however, this is reduced to less than $0.001t$ if quantum number projections are applied. In the following study, however, we do not apply quantum number projections due to numerical cost, because, as we demonstrate in Appendix \ref{s:projections}, the effect of quantum number projections on other physical quantities is relatively small.

Away from half-filling, exact results are not available for comparison, however, the method achieves close agreement with the recent results from other methods at 1/8 doping at $U/t=8$, a point which was recently intensively studied \cite{zheng_stripe_2017}. 

\section{Phase diagram}
\label{s:phase_diagram}
\begin{figure}
\includegraphics[width=0.5\textwidth]{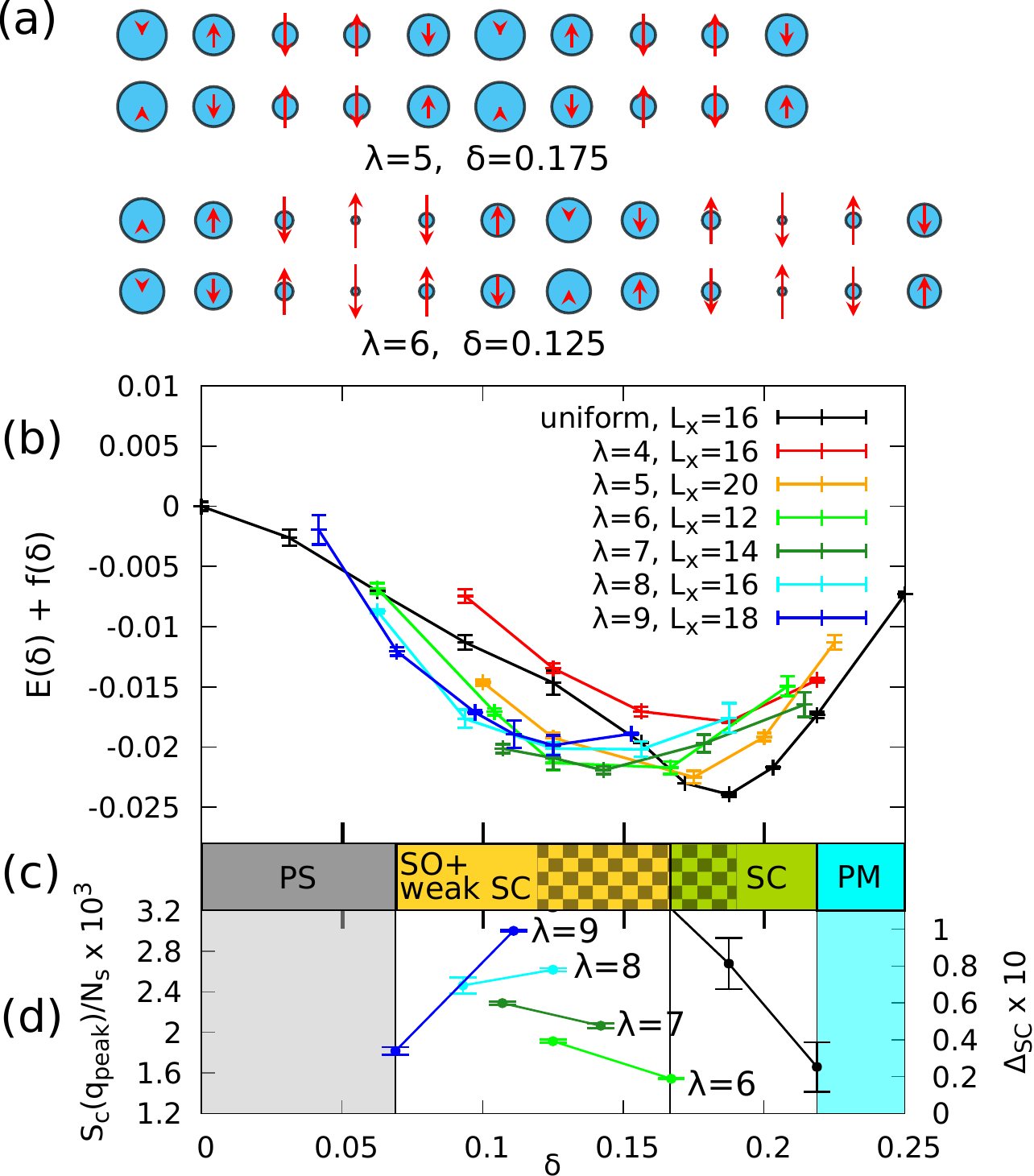}
    \caption{(a) Examples of hole density and $\langle S_z \rangle$ in a $2\lambda\times 2$ cell for an odd and even period stripe obtained by optimization at $U/t=10$. Hole density is proportional to the diameter of the circle, while $\langle S_z \rangle$ is proportional to the length of the arrow. (b) Energies of stripe and uniform states per site in units of $t$ as a function of doping for $L_y=8$ systems ($U/t=10$). $L_x$ is indicated in the legend. As finite-size effects are found to be small (see Appendix \ref{s:finite}) we note that the result here essentially represents the thermodynamic limit.  
    A linear function $f(\delta)= 1.835\delta+0.4304$ has been added to the energy in (b) to improve visibility. (c) Ground-state phase diagram with phase separation (PS),
stripe 
 spin and charge
order coexisting with weak $d$-wave superconductivity (SO $+$ weak SC),
charge uniform $d$-wave SC (SC) and charge uniform paramagnetic normal (PM) regions 
indicated. Checkerboard region indicates a possible region of phase separation between SO and uniform SC phases. (d) Thermodynamic limit extrapolations of order parameters of the ground state in different regions: peak of charge structure factor is plotted for the SO region, and $\Delta_{\rm SC}$ is plotted in the PM region. Charge structure of several different stripe periods is plotted simultaneously as different stripe periods are near degenerate.}
\label{f:phase_diagram}
\end{figure}

The energies of charge uniform and various stripe states, with 
charge
periods ranging from 4 to 9, were calculated as a function of doping as shown in Fig. \ref{f:phase_diagram}(b).
 The strength of the on-site Coulomb repulsion was fixed to $U/t=10$, which is close to {\it ab initio} estimates for cuprates~\cite{hirayamaab2017}. 
Each point represents the ground state of an $L_y=8$ system obtained by variance extrapolation. While only fixed system sizes are displayed, we have carried out a careful analysis of size effects, finding that further increasing the system size makes only a small change to energies, and does not change the essential features in the diagram. Full details of finite-size checks are included in 
Appendix \ref{s:finite}.

Firstly, we remark that for values of doping up to $\delta=0.25$, 
many different orders are severely competing and the ground-state phase is determined by subtle energy balance: 
the energy differences between different stripe states and uniform states is small, typically less than $0.01t$ ($\sim 50$ K in cuprate energy scale~\cite{hirayamaab2017}). 
This is smaller by about a factor of two compared to the energy difference observed using the variational mVMC energies in Ref.~\onlinecite{ido_competition_2018}. 
This tendency indicates that improving the accuracy of the wave function results in a larger energy reduction for the charge uniform (superconducting as will be clarified later) state than the stripe ordered state. This was also observed in iPEPS~\cite{corboz_competing_2014}. Relative stabilization of the charge uniform state for a better wave function may be understood by the fact that the charge uniform metal and $d$-wave superconducting off-diagonal ordered states are subject to larger quantum fluctuations than simple charge/spin symmetry-broken states and require more sophisticated wave functions. In contrast, the stripe-type diagonal symmetry-broken states can be represented already by the mean-field level relatively well and the sophisticated wave functions do not improve the energy as much as the charge uniform states.

The ground-state phase diagram from $\delta=0$ to $\delta=0.25$
is shown in in Fig. \ref{f:phase_diagram}(c) and the relevant order parameters determined by TL extrapolation are shown in Fig. \ref{f:phase_diagram}(d).

For large values of doping $0.17 \lesssim \delta \lesssim 0.25 $, the ground state is homogeneous. 
The uniform ground state was not observed in the less accurate mVMC study for this doping range \cite{ido_competition_2018} and could provide insight into the mechanism for high-temperature superconductivity in cuprates. 
The superconductivity is seen in the region $0.17 \lesssim \delta \lesssim 0.22$. The staggered antiferromagnetic order, seen at low doping of uniform state, happens to disappear continuously at $\delta=0.17$ as we discuss later.

In the region $0.07 \lesssim \delta \lesssim 0.17$, the ground state is stripe ordered with very small energy differences between stripes states of different periods. The preferred stripe period decreases with increasing doping, as the mean distance between the holes decreases. The extent of this region agrees well with experiments on cuprates, in which charge inhomogeneities have been observed in the doping range $0.05\lesssim\delta\lesssim 0.2$ \cite{tranquada_evidence_1995, tranquada_coexistence_1997, yamada_doping_1998, hucker_enhanced_2013, fink_phase_2011, ghiringhelli_long-range_2012, tabis_charge_2014, comin_broken_2015, comin_symmetry_2015, forgan_microscopic_2015, peng_direct_2016, campi_inhomogeneity_2015, mesaros_commensurate_2016}.

Although we have used a larger coupling strength of $U/t=10$, our results qualitatively agree with those of Ref.~\onlinecite{zheng_stripe_2017} with $U/t=8$ at doping point $\delta=0.125$. We observe near degeneracy of stripes with periods from 5 to 8, with period 4 stripe and uniform states being about $0.01t$ higher in energy. Although in agreement with other numerical methods, this deviates from the experimentally observed period of around 4 at $\delta\approx 1/8$ in La-based cuprates \cite{tranquada_evidence_1995, tranquada_coexistence_1997}. As was recently shown in Ref.~\onlinecite{ido_competition_2018}, this discrepancy can be explained by the absence of next-nearest hopping in the simple square-lattice model. 

In the region $\delta \lesssim0.07$, the energy follows a slightly downward concave path as the ground state transitions from uniform state to a long period stripe.
This suggests that phase separation occurs between the Mott insulator at half-filling and a long period stripe. 
We have limited our calculations to a maximum charge period of 9. Longer charge periods were not considered due to their prohibitive computational cost. Since the period of the stripe in the ground state increases with decreasing carrier doping it is conceivable that longer period stripe may appear for $\delta <0.07$, which fills the phase separation region. This is left for future studies.

\section{Charge and spin correlations}
We now discuss the physical properties of the ground state in more detail. We have found that applying first-step Lanczos to the variational wave function makes little change to spin, charge and pairing correlations, in agreement with Ref.~\onlinecite{{ido_competition_2018}} (see Appendix D of that work for more details). Therefore, unless otherwise specified, physical quantities in the following discussion are obtained using the variational wave function which includes the tensor network correlation factor but without first-step Lanczos applied. 

In order to quantify charge and spin correlations, we have calculated structure factors as a function of doping. The spin structure factor is defined as 
\begin{equation}
    S_s(\bm{q})=\fr{3N_s}\sum_{i,j}\langle \bm{S}_i\cdot \bm{S}_j \rangle e^{i\bm{q}\cdot(\bm{r}_i-\bm{r}_j)}\,,
\end{equation}
where $\bm{r}=(r_x, r_y)$ is the site position. The charge structure factor is defined as 
\begin{equation}
    S_c(\bm{q})=\fr{N_s}\sum_{i,j}\langle (n_i - n)(n_j - n ) \rangle e^{-i \bm{q}\cdot (\bm{r}_i - \bm{r}_j)}\,,
\end{equation}
where $n=N/N_s$. 

\begin{figure}
    \includegraphics[width=0.40\textwidth]{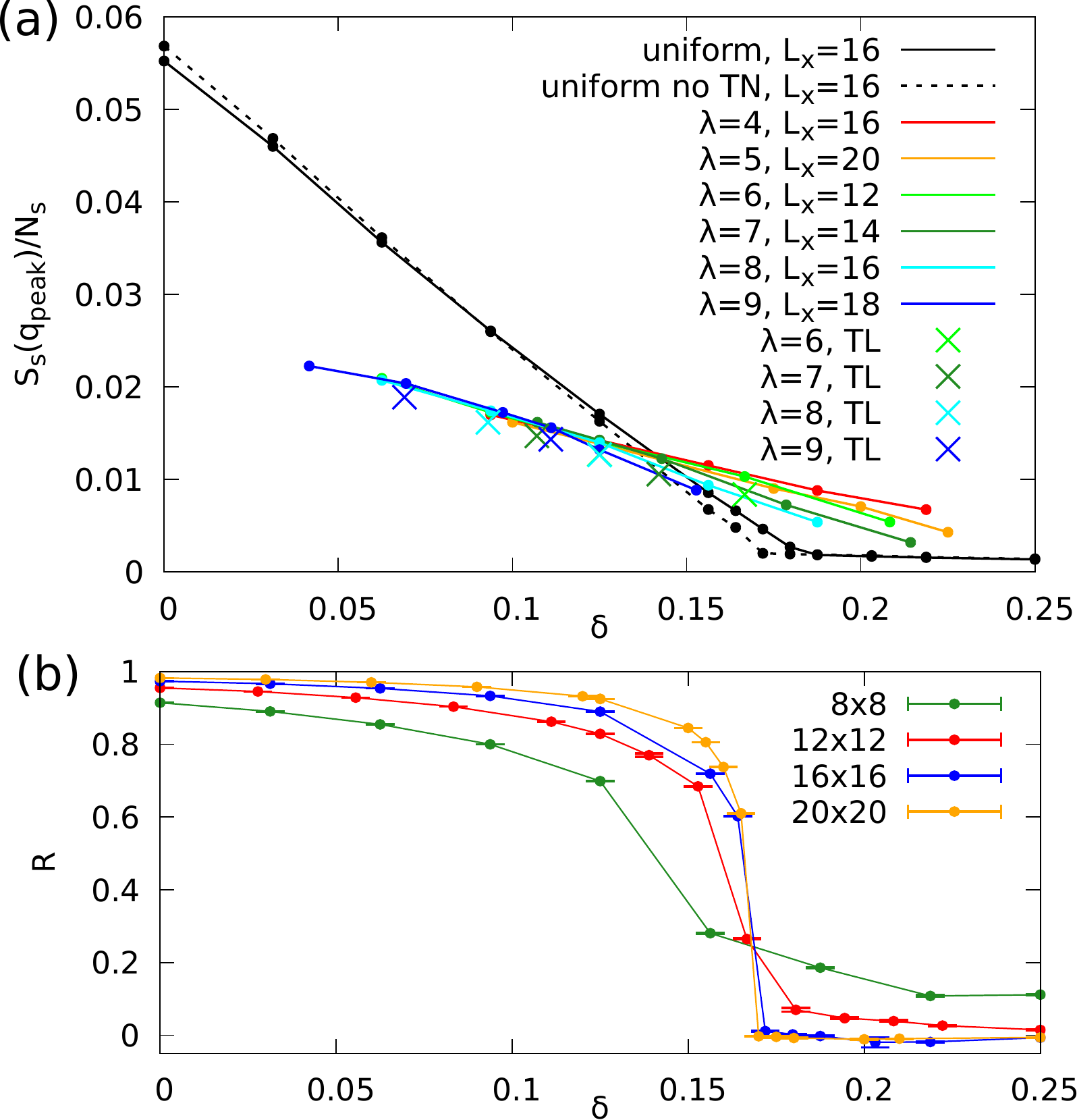}

    \caption{(a) Peak values of spin structure factor and (b) correlation ratio as a function of doping $\delta$. In (a) dots are from finite-size ($L_y=16$) calculations,  while cross symbols are obtained as values after the TL extrapolation. (a) shows results with the tensor network correlation factor, while (b) is calculated without it due to computational cost.}
    \label{f:structure}
\end{figure}

The peak values of $S_s({\bm q})$ and $S_c({\bm q})$ for different states as a function of doping are shown in Figs. \ref{f:phase_diagram}(d) and \ref{f:structure}(a), respectively. For $S_s({\bm q})$, results for both finite-size systems and TL extrapolated values are shown. Details of the TL extrapolation are included in 
Appendix \ref{s:phys_size_extrap}.

In the stripe region we obtain nonzero TL values for $S_s({\bm q})$ and $S_c({\bm q})$, which confirms that stripe order persists in the TL. 

For the charge uniform state, a transition from an anti-ferromagnetic phase to a paramagnetic phase is observed at $\delta_c \approx 0.17$, where the continuous reduction of $S_s$ to zero is consistent with a continuous or very weak first-order transition. We establish the robustness of anti-ferromagnetic order below $\delta_c$ using the correlation ratio \cite{kaul_spin_2015, pujari_interaction-induced_2016}, defined as 
\begin{equation}
    R= 1-\frac{S_s({\bm q}_{\rm peak}+\delta {\bm q})}{S_s({\bm q}_{\rm peak})}
\end{equation}
where ${\bm q}_{\rm peak}$ is the point where $S_s$ takes its maximum value and ${\bm q}_{\rm peak}+\delta {\bm q}$ the  closest neighboring point. As the system size tends to infinity,  $R\rightarrow 1$ in an ordered phase, $R\rightarrow 0$ in a disordered phase and $R$ will be a
constant independent of $N_s$ at a critical point. The correlation ratio is plotted as a function of doping in Fig. \ref{f:structure}(c), in which the antiferromagnetic quantum critical point is again suggested at $\delta\approx0.17$. Although the weak first-order transition is not excluded, the transition point $\delta=0.17$ is well determined.

\section{Superconducting order}

Here we investigate superconducting correlations in various low energy states of the Hubbard model. We quantify superconductivity with the  $d_{x^2-y^2}$-wave superconducting correlation function 
\begin{equation}
P_d(\bm{r})=\fr{2N_s}\sum_{\bm{r_i}} \langle \Delta_d^\dag(\bm{r}_i)\Delta_d(\bm{r_i+r})+\Delta_d(\bm{r}_i)\Delta_d^\dag(\bm{r_i+r})\rangle\,,
\label{e:sc}
\end{equation}
where
\begin{equation}
    \Delta_d(\bm{r_i})=\rt{2}\sum_{\bm{r}} f_d(\bm{r})(c_{\bm{r_i}\uparrow}c_{\bm{r_i+r}\downarrow} - c_{\bm{r_i}\downarrow}c_{\bm{r_i+r}\uparrow})\,,
\end{equation}
$f_{d}(\bm{r})=\delta_{r_y, 0}(\delta_{r_x, 1}+\delta_{r_x, -1})-\delta_{r_x, 0}(\delta_{r_y, 1}+\delta_{r_y, -1})$ is the $d_{x^2-y^2}$-wave form factor and $\delta_{i,j}$ is the Kronecker delta. In our calculations we define the superconducting order parameter as $\Delta_{\rm SC}=\sqrt{P_d^\infty}$, where $P_d^\infty=\fr{|A|}\sum_{\bm{r} \in A} P_d(\bm{r})$ is the long-range correlation function, which is averaged over a set $A$ of sufficiently large displacements to smooth out fluctuations. For stripe states, rather than averaging over all sites $\bm{r}_i$ in Eq. \eqref{e:sc}, we measure correlations along hole rich stripes, which typically have stronger correlations. 

The main question that we seek to answer is whether 
stable superconducting phase exists in the Hubbard model or not.
In the charge uniform state, the doping dependence of superconducting order has a dome shape with a maximum at around $\delta=0.125$. The dependence for a finite-sized system was found to be qualitatively identical to that shown in Ref.~\onlinecite{ido_competition_2018}, so we have not included the figure here.
After the TL extrapolation (see details in Appendix \ref{s:phys_size_extrap}), the superconducting order survives up to $\delta \approx 0.22$, as shown in Fig. \ref{f:phase_diagram}(d). 
 Below $\delta \approx 0.17$, superconducting and antiferromagnetic orders coexist.

We also have indications for nonzero $\Delta_{\rm SC}$ in certain stripe states, however, it is much smaller than the uniform state, and does not have systematic size dependence, making TL extrapolation difficult. A rough extrapolation was performed using the mVMC wave function without the TN. We found $\Delta_{\rm SC}=0.052\pm 0.023$ in a $\lambda=7$ stripe at $\delta=0.107$ and $\Delta_{\rm SC}=0.045\pm 0.018$ in a $\lambda=6$ stripe at $\delta=0.125$. We also found nonzero $\Delta_{\rm SC}$ in $\lambda=8$ and $\lambda=9$ stripes for finite-size systems, with similar magnitude to the $\lambda=6$ and $\lambda=7$ stripes. However, due to computational cost, we were unable to estimate TL values for these systems.

\label{s:orders}

It appears that $\Delta_{\rm SC}$ is a challenging quantity to estimate, with different methods obtaining different values in stripe and uniform states \cite{zheng_stripe_2017}. Nevertheless, in agreement with DMET and iPEPS in Ref.~\onlinecite{zheng_stripe_2017}, we have shown that $\Delta_{\rm SC}$ is more robust in charge uniform states than stripe states. The uniform state is the ground state at $\delta \gtrsim 0.17$ and $\Delta_{\rm SC}$ decreases with doping in this region, as seen in Fig. \ref{f:phase_diagram}(d). At lower doping $\delta \lesssim 0.17$, this state competes with the stripe states, which are less superconducting. Superconductivity in the ground state therefore appears optimal at around $\delta\approx 0.17$, which is in good agreement with the optimal doping in cuprates of around $\delta\approx 0.16$~\cite{uchida_recent_1993}.

Figure \ref{f:phase_diagram} hints at an intriguing possibility of phase separation between the superconducting state at $\delta\approx 0.19$ and the stripe state at $\delta\approx 0.12$. This could explain the  gradual decrease in critical temperature away from optimal doping in this region. 
In reality, this region may remain as the charge inhomogeneous phase with the volume fraction of the superconducting states $R_s=(\delta-0.12)/(0.19-0.12)$ and $R_c=1-R_s$ for the stripe states.  Such reduction of the volume fraction may alter the critical temperature as in the granular superconductivity \cite{lang_imaging_2002, imry_destruction_1981}.
 However, the existence of phase separation depends on the precise shape of the uniform state's energy curve. Precise numerical calculations are challenging in this region due to the antiferromagnetic quantum critical point at around $\delta\approx0.17$. We leave the detailed study of this feature to future work. 

\section{Conclusion}
\label{s:conclusions}
In this work, we have performed a systematic study of the hole-doped Hubbard model on a square lattice at strong coupling $U/t=10$, focusing on how the ground state properties vary as a function of doping. We have employed a variational wave function which combines a Pfaffian with a tensor network to efficiently represent the different types of entanglement likely to be present.  
Our method is substantially more accurate than the previous mVMC study \cite{ido_competition_2018}. Our improved method has enabled us to uncover a charge uniform and strong $d$-wave superconducting phase near $\delta\approx 0.2$ sandwiched by the paramagnetic normal metal phase for $\delta\gtrsim0.22$ and stripe phase with doping-dependent periodicity for $\delta\lesssim0.17$.  This region was formerly identified as the stripe ordered ground state with either period 5 or 8 \cite{ido_competition_2018}.  However, the present, more accurate method has exposed the existence of a small window with the superconducting order in the so-called overdoped region.
These phases are severely competing within the energy scale of $0.01t$ for all the doping $\delta>0$ studied. The possible phase separation suggests a coexistence of stripe and superconducting domains roughly for $0.12\lesssim\delta\lesssim0.19$. Possible weak superconductivity is also found in the stripe ground states at low doping.
It is remarkable that the simplest Hubbard model studied here qualitatively reproduces the basic experimental phase diagram of the cuprates with various similarities. 

However, a very wide region ($0.07\lesssim\delta\lesssim0.17$) of stripe long-range order with strongly suppressed (or vanishing) $d$-wave superconducting order is required to be critically compared in the future with the experimental phase diagrams for most of cuprate compounds dominated by the $d$-wave superconductivity at lowest temperatures.
An interesting direction for future study would be to observe how physical properties change when additional terms are added to the Hamiltonian to more realistically describe the physics of cuprates, for instance, terms obtained from {\it ab initio} studies \cite{hirayamaab2017}. Starting from an accurate {\it ab initio} effective Hamiltonian for the cuprates, its reliable solution with detailed and quantitative comparison with the cuprates will open the materials design for further enhancing superconductivity. For instance, the enhancement of superconductivity due to laser irradiation \cite{ido_correlation-induced_2017} has recently been investigated using a similar VMC technique.

\acknowledgements
The present work was supported by JSPS KAKENHI
(Grants Nos. 16H06345 and 17K14336) from Ministry of Education, Culture, Sports, Science and Technology (MEXT), Japan.
This research was also supportd by MEXT as
``Priority Issue on Post-K computer" [Creation of New Functional Devices and High-Performance Materials
to Support Next-Generation Industries (CDMSI)] with the project supported by RIKEN Advanced Institute for Computational Science
(AICS) through HPCI System Research Project (Grants No.
hp170263 and No. hp180170).
The authors thank the 
Supercomputer Center, the Institute for Solid State Physics, 
the University of Tokyo for the facilities. 
\appendix

\section{Detailed description of method}
\label{s:method}
Here we provide a more detailed description of the method. The variational wave function in Eq \eqref{e:wavefunction}, has three components: a Pfaffian $\psi_{\rm pair}(x)$, correlation factors $\mathcal{P}(x)$, and a fat tree tensor network $\mathcal{M}(x)$.
The Pfaffian term represents a pair-product wave function, defined as 
\begin{equation}
    \ket{\psi_{\rm pair}}=\left(\sum_{i,j} f_{ij} c_{i\uparrow}^\dag c_{j\downarrow}^\dag\right)^{N/2}\ket{0}\,,
\end{equation}
where $f_{ij}$ are variational parameters. This wave function can exactly represent various types of states typically found in strongly correlated quantum systems, including mean-field superconducting,  charge-ordered and antiferromagnetic states, resonating valence bond solid states and many others. For any real space configuration $\ket{x}$, the overlap $\psi_{\rm pair}(x):=\braket{x}{\psi_{\rm pair}}$ is the Pfaffian of a matrix, which can be computed efficiently. 

The correlation factors are given by $\mathcal{P}=\mathcal{P}_J\mathcal{P}_G\mathcal{P}_{d-h}$ where $\mathcal{P}_G=\exp \left[ -\fr{2} \sum g_{i} n_{i\uparrow}n_{i\downarrow} \right]$ is the Gutzwiller factor, $\mathcal{P}_J=\exp \left[ \fr{2} \sum v_{ij} n_{i}n_{j} \right]$ is the Jastrow factor, and $\mathcal{P}_{d-h} = \exp\left[ -\sum_{m=0}^4 \alpha_{(m)} \sum \xi_{(m)} \right]$
is the doublon-holon factor, $n_i=n_{i\uparrow}+n_{i\downarrow}$ is the number of electrons at site $i$, and $\xi_{(m)}$ is 1 when a doublon (holon) exists at site $i$, with $m$ holons (doublons) at nearest-neighbor sites. These factors are all diagonal in the real-space configuration basis. The variational parameters $g_i$, $v_{ij}$, and $\alpha_{(m)}$ are optimized by the method. The Gutzwiller factor can take into account local correlation effects, while the Jastrow factor and doublon holon factors can take into account longer range correlations, which are particularly important in describing correlations in Mott insulators \cite{capello_variational_2005}. 

The combination of the tensor network $\mathcal{M}(x)$ and $\mathcal{P}\ket{\psi_{\rm pair}}$ was first introduced in Ref.~\onlinecite{zhao_variational_2017-1}, in which a full description is available. In this tensor network, entangled plaquettes of four sites are coupled via a binary tree tensor network. Such a tensor network can flexibly represent types of area law entanglement that are not captured by $\mathcal{P}$ or $\psi_{\rm pair}$. A parameter $D$ specifies the bond dimension, with larger $D$ resulting in a larger number of parameters, and a more accurate variational wave function. Evaluation of $\mathcal{M}(x)$ for any given real space configuration $x$ consists of the contraction of a binary tree tensor network, which can be performed exactly and efficiently in time $\mathcal{O}(D^3 N_s)$. In most of our calculations we set $D=2$. While it is possible to obtain a more accurate wave function by increasing $D$, we found it more efficient to apply first order Lanczos to the $D=2$ wave function than to increase $D$.

In variational Monte Carlo, expectation values of local observables are estimated by sampling over the probability distribution $p(x)=\braket{\psi}{x}\braket{x}{\psi}/\braket{\psi}{\psi}$ using Markov chain Monte Carlo. This is possible for the variational wave function in Eq. \eqref{e:wavefunction}, because $\mathcal{P}(x)$, $\mathcal{M}(x)$ and $\psi_{\rm pair}(x)$ can be efficiently calculated for any $x$. The doping $\delta$ is fixed by sampling only over configurations $\ket{x}$ with a fixed number of electrons $N$. 

Given that derivatives of the wave function with respect to variational parameters can be calculated, the variational parameters can be optimized with respect to the energy using the stochastic reconfiguration (SR) method \cite{sorella_generalized_2001}.
We use a version of SR that employs conjugate gradient to avoid constructing the SR matrix explicitly, allowing a large number of parameters to be simultaneously optimized \cite{neuscamman_optimizing_2012}. 

Calculating the energy requires time complexity scaled by 
$\mathcal{O}(n_s(N_s^3+D^3 N_s\log_2(N_s)))$, where $n_s$ is the number of samples. The first term comes from the calculation of the Pfaffian, while the second term comes from the tensor network contraction. The optimization has a time complexity of $\mathcal{O}(n_s n_p n_{\rm iter})$, where $n_{\rm iter}$ is the number of iterations in solving the SR equation with conjugate gradient and $n_p$ is the number of parameters, which scales as $\mathcal{O}(N_s^2+N_s D^3)$ if a full unit cell is used.

\begin{figure}
\includegraphics[width=0.45\textwidth]{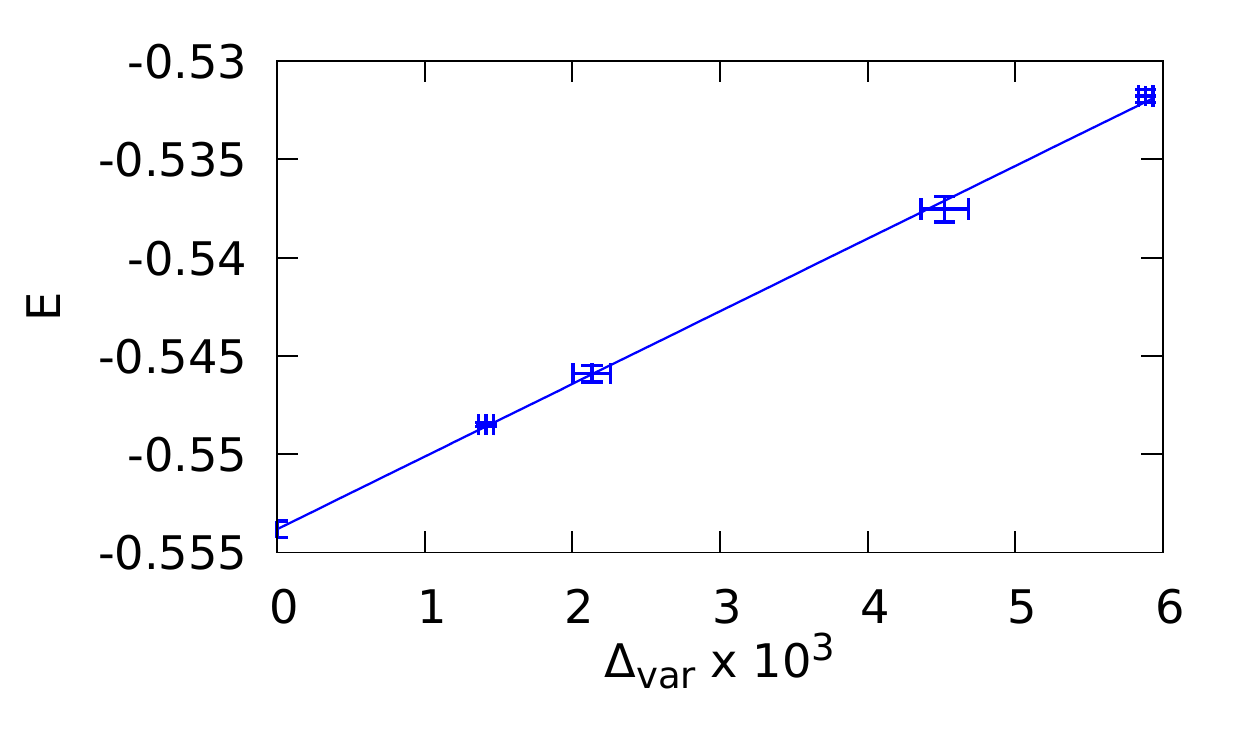}
    \caption{Variance extrapolation of the energy per site in units of $t$ at $\delta=0.0625$ on a $16\times8$ system at $U/t=10$. Points on this plot from highest to lowest energy are the VMC wave function without TN, VMC wave function with TN and $D=2$, VMC wave function without TN with first-step Lanczos applied and VMC wave function with $D=2$ TN and first-step Lanczos.}
\label{f:varext}
\end{figure}

The ground state of a finite-sized system has a number of symmetries which can be exploited to further improve the energy. Projections that restore translation, $SU(2)$ and rotational $C_4$ symmetry, denoted respectively by $\mathcal{L}^K$, $\mathcal{L}^S$ and $\mathcal{L}^{C_4}$, can be applied to the ground state by modifying the $\mathcal{M}(x)$ and $\psi_{\rm pair}(x)$ terms in Eq. \eqref{e:wavefunction}. Details of how such quantum number projections~\cite{ring_nuclear_2004} are implemented can be found in Ref.~\onlinecite{tahara_variational_2008}. Since these projections are computationally expensive and result in a relatively small improvement for large system sizes, they are used only in certain cases. As we show in 
Sec. \ref{s:projections}, 
these projections also have a negligible effect on other physical quantities, such as spin and charge correlations. 

\section{Finite-size effect and thermodynamic limit}
\label{s:finite}
To ensure that the quantities calculated are representative of the TL values, we have performed a number of finite-size checks. Regarding the energy, our basic observation is that under our calculation conditions, and for sufficiently large systems, the energies obtained by variance extrapolation are largely insensitive to system size.
Since we employ the periodic boundary condition in the $y$ direction while antiperiodic in the $x$ direction, the two directions are not equivalent.  Then we discuss below the $L_x$ and $L_y$ dependencies separately. Of course, in the limit of both $L_x,L_y \rightarrow \infty$, the unique TL values should be recovered.

\subsection{Size dependence of energy}
\begin{figure}[t]
\includegraphics[width=0.35\textwidth]{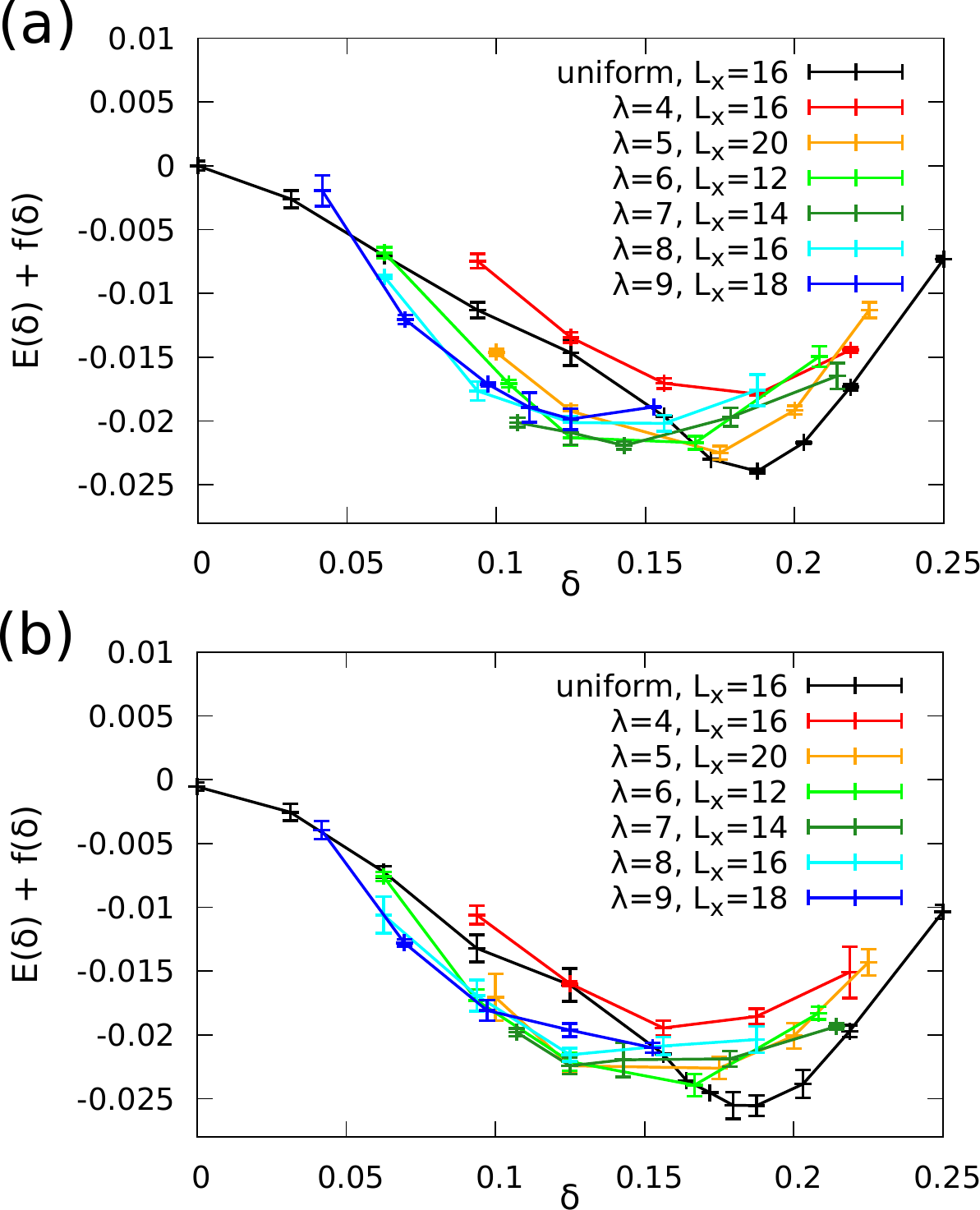}
    \caption{Comparison of energy per site (in units of $t$) vs doping plot at $U/t=10$ for two different system sizes: (a) with $L_y=8$ and (b) with $L_y=16$. Essential features of the diagram appear unaffected by increasing size beyond $L_y=8$ . A linear function $f(\delta)= 1.835\delta+0.4304$ has been added to the energy to improve visibility. }
\label{f:phase_diagram_sizecompare}
\end{figure}

Let us first discuss $L_y$ dependence. Although we have presented only the $L_y=8$ plot in Fig. \ref{f:phase_diagram} in the main text, we have also calculated energies for different sized systems. The entire plot for $L_y=16$ is shown below the $L_y=8$ plot in Fig. \ref{f:phase_diagram_sizecompare}. All essential features of the diagram and even the energy differences of various stripe orders and uniform state are preserved within the range of error bars.
The insensitivity supports that the results obtained by the size $L_y=8$ is close to the TL results. 

We adopted $L_y=8$ because we can perform more stable energy-variance extrapolations for the following reason: 
although the energy is relatively weakly dependent on the energy variance, the improvement of the energy variance obtained from taking first-step Lanczos and employing larger number of tensor dimension in the tensor network part tends to decrease when the system size becomes larger.  Namely, the energy variance stays relatively higher for larger systems for the same level of Lanczos and tensor-network treatment and remains more distant from the limit of the zero-energy-variance extrapolation.
Then the extrapolation causes larger errors in the extrapolated energy.
This explains why the error bars on the $L_y=16$ plot are relatively larger than on the $L_y=8$ plot in Fig. \ref{f:phase_diagram_sizecompare}. 
Since the energies at $L_y=8$ lie more or less within the increased error-bar range of $L_y=16$, the size extrapolation is not meaningful. However, the weak system size dependence of the energies suggests that the results at $L_y=8$ are already close to the TL energies even for the relative order of energies for different periods of stripes despite their severe competitions. Therefore, we show the result for $L_y=8$ as a good estimate of the TL phase diagram.

We next discuss the $L_x$ dependence.  At certain doping points, system size was also extended in the $L_x$ direction, and we observe similar  size insensitivity. Energies obtained with different system sizes at $U/t=10$ are shown in Table \ref{t:energy_size}.
\begin{table}[t]
\begin{tabular}{|c|c|}
\hline
    System size ($L_x\times L_y$)& Energy per site\\
\hline
\hline
\multicolumn{2}{|c|}{Stripe $\lambda=5$, $\delta=0.2$ }\\
\hline
$10\times 8 $& -0.8195(6)\\
$10\times 16$& -0.8180(8)\\
$20\times 8 $& -0.8165(3)\\
$20\times 16$& -0.8175(9)\\
\hline
\multicolumn{2}{|c|}{Stripe $\lambda=6$, $\delta=0.167$ }\\
\hline
$12\times 8 $& -0.7579(5)\\
$24\times 8 $& -0.7589(2)\\
$12\times 16$& -0.7601(9)\\
\hline
\multicolumn{2}{|c|}{Stripe $\lambda=7$, $\delta=0.143$ }\\
\hline
$14\times8 $& -0.7144(3)\\
$28\times8 $& -0.7150(3)\\
$14\times16$& -0.716(1) \\
\hline
\multicolumn{2}{|c|}{Stripe $\lambda=8$, $\delta=0.125$ }\\
\hline
$16\times8 $& -0.6798(5)\\
$16\times16$& -0.6813(5)\\
$32\times8 $& -0.681(1) \\
\hline
\multicolumn{2}{|c|}{Uniform, $\delta=0.125$}\\
\hline
$16\times8$&-0.674(1)\\
$16\times16$&-0.676(1)\\
$32\times8$&-0.677(2)\\

\hline
\end{tabular}
\caption{Energies per site in units of $t$ obtained for different system sizes at $U/t=10$.}
\label{t:energy_size}
\end{table}
The energies depend very little on the system size again and the energies of different sizes are indistinguishable within the error bars, if $L_x$ is larger than 10.
Even when we perform the energy extrapolation using the variance, for both stripe and uniform states, 
doubling the system size results in only a slight decrease in extrapolated energy of approximately $\sim 0.001t$ to $0.002t$ and the relative energy difference of various competing orders hardly changes.  Therefore in the main text we use $12\le L_x\le20$.

\subsection{Size extrapolation of physical quantities}
\label{s:phys_size_extrap}
In this section, we describe how TL values are calculated for charge, spin and superconducting correlation functions. 
As mentioned in the main text, applying first-step Lanczos resulted in little change to physical quantities except for the energy. Physical quantities also changed little when the tensor-network bond dimension was increased beyond $D=2$. The physical quantities described below are therefore calculated using the variational wave function without first-step Lanczos applied, and with the tensor-network bond dimension set to $D=2$.

\subsubsection{Spin and charge structure factors and orders}
\begin{figure}[t]
\includegraphics[width=0.48\textwidth]{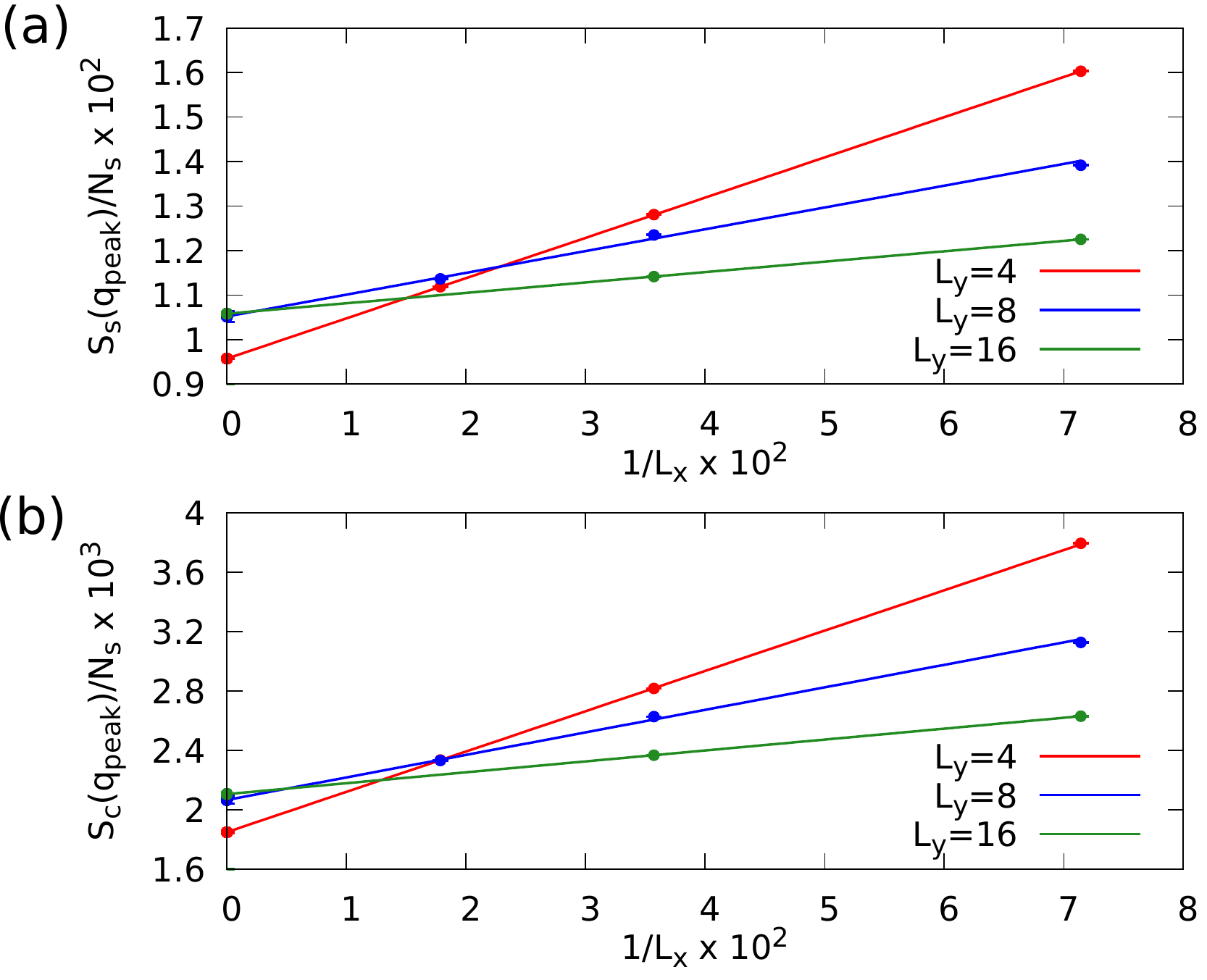}
    \caption{System-size extrapolation for (a) spin structure factor peak and (b) charge structure factor peak for a $\lambda=7$ stripe at $\delta=0.142$ and $U/t=10$. Error bars are smaller than marker sizes. Extrapolated values for infinitely long $L_y=8$ systems and $L_y=16$ systems are essentially indistinguishable. For these system sizes, the one and two dimensional extrapolations with respect to $N_s$ are nearly identical, therefore we have taken the one dimensional extrapolation with fixed $L_y=8$ as the TL extrapolated values of $S_s$ and $S_c$.}
\label{f:sf_sizeextrap}
\end{figure}

 In order to obtain the TL values of structure factors, we first fix $L_y$ and extrapolate to infinite $L_x$, giving the structure factor of an infinitely long strip. This is shown for an $\lambda=7$ stripe in Fig. \ref{f:sf_sizeextrap} for $L_y=4,8,16$. We observed that infinite $L_x$-extrapolated values for spin and charge structure factor peaks was the same when $L_y$ was set to 8 or 16 implying that an infinitely long $L_y=8$ system is already representative of the TL for spin and charge structure factors. 
We note that taking the limit as $L_x\rightarrow\infty$ with fixed $L_y$ is expected to yeild a Tomonaga-Luttinger liquid with vanishing long-range order, i.e. $S_s/N_s \rightarrow 0$. While this is the expected behavior when $L_x\gg L_y$, in our calculations, where $L_x \sim L_y$ the value of $S_s$ and $S_c$ appears largely independent of system shape when $N_s$ is fixed because the employed system sizes here are essentially in the two-dimensional scaling region and the characteristic one-dimensional size dependence is expected to appear at much larger aspect ratio. For instance, the peak values of $S_s$ are close (within $0.5\%$) for a $28\times16$ system and a $56\times 8$ system for the $\lambda=7$ stripe. Thus, for the system sizes used in these calculations, the one and two dimensional extrapolations  with respect to $N_s$ are comparable, which justifies the use of one-dimensional extrapolations.   
TL values in Fig. \ref{f:phase_diagram} and Fig. \ref{f:structure} were therefore obtained by extrapolation of an $L_y=8$ strip.  

\subsubsection{Superconducting order}
\begin{figure}
\includegraphics[width=0.48\textwidth]{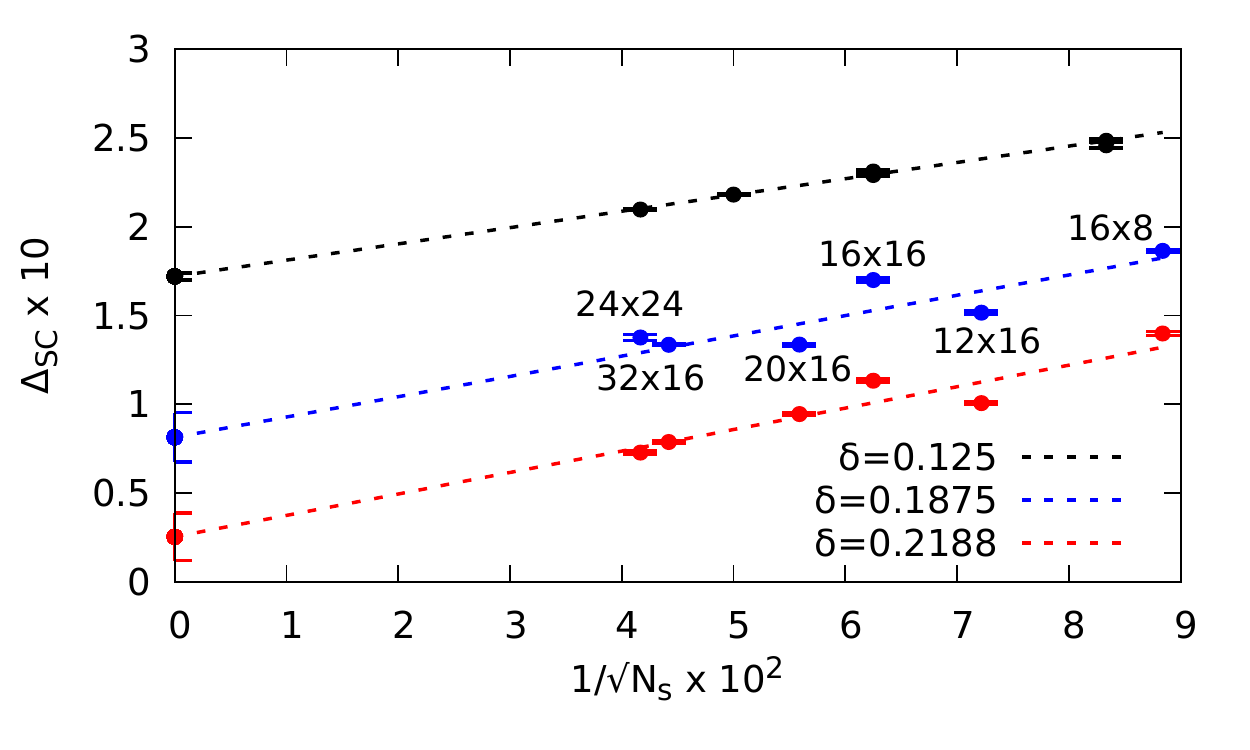}
\caption{Size extrapolation of $\Delta_{\rm SC}$ in charge uniform states. Calculations were performed at $U/t=10$ and the system size is varied. At $\delta=0.125$ only square systems with $L_x=L_y$ were used, however at other dopings, non-square lattices were needed (since square systems could not support the specified doping for an integer number of electron pairs). The system dimension is labeled for $\delta=0.1875$. The same dimensions were used for $\delta=0.2188$. For $\delta=1875$ and $\delta=0.2188$, values obtained with antiperiodic-periodic and periodic-periodic boundary conditions were averaged to reduce finite-size effects. The tensor-network bond dimension was fixed to $D=2$, as we observed little change with increasing $D$ beyond this.}
\label{f:uniform_sc_extrap}
\end{figure}

We now provide details on how TL values of $\Delta_{\rm SC}$ were calculated.  In Fig. \ref{f:uniform_sc_extrap} we plot $\Delta_{\rm SC}$ vs. system size for the charge uniform state at three dopings $\delta=0.125,0.1875, 0.2188$. For $\delta=0.125$, four sizes were considered: $12\times12$, $16\times16$, $20\times20$ and $24\times24$. At this doping, superconducting order $\Delta_{\rm SC}$ scales linearly with the inverse linear dimension $1/\sqrt{N_s}$, and extrapolation to the TL yields a large nonzero value of $\Delta_{\rm SC}=0.172(2)$.

For the other values of doping, different system sizes, including nonsquare systems, had to be used (since the method requires an integer number of electron pairs, a given doping can only be supported on certain  system sizes). The sizes used in both $\delta=0.1875$ and $\delta=0.2188$ were $16\times 8$, $12\times 16$, $16\times 16$, $20\times 16$, $32\times16$ and $24\times 24$. The extrapolation was performed with respect to $\sqrt{N_s}$ which equals $L$ for square systems. Some fluctuation in system size was observed, which was reduced by averaging over results obtained by periodic-periodic and antiperiodic-periodic boundary conditions.  We observed that $\Delta_{\rm SC}$ remained robust at $\delta=0.1875$, although somewhat smaller than at $\delta=0.125$. Superconductivity decreased rapidly to near zero at $\delta=0.2188$.

We have also estimated $\Delta_{\rm SC}$ in certain stripe states. We remark that calculating superconductivity in stripe states is challenging and different methods do not agree on the value of $\Delta_{\rm SC}$. For example, in Ref.~\onlinecite{zheng_stripe_2017}, iPEPS found nonzero $\Delta_{\rm SC}$ in $\lambda=5$ and $\lambda=7$ stripes at $\delta=0.125$ and $U/t=8$, while DMET only found nonzero $\Delta_{\rm SC}$ in $\lambda=9$ and in a metastable excited stripe state with $\lambda=5$.

\begin{figure}
\includegraphics[width=0.48\textwidth]{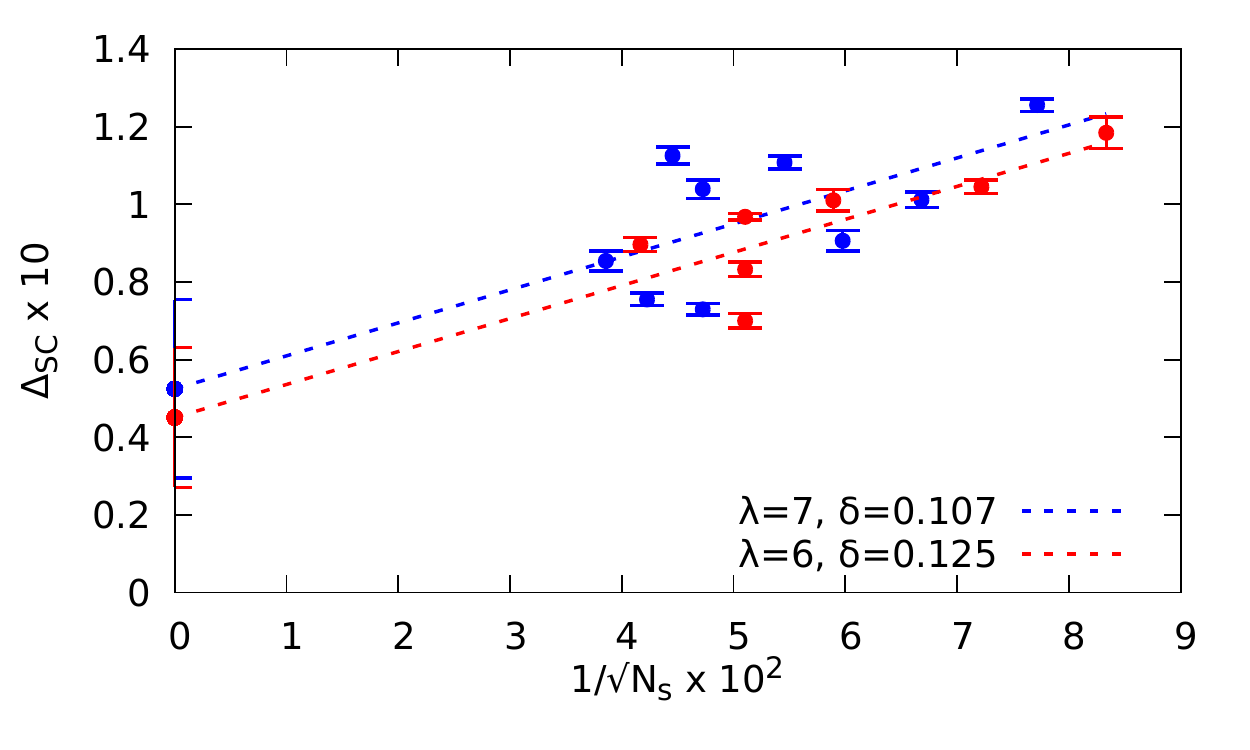}
\caption{Size dependence of $\Delta_{\rm SC}$ in stripe states at $U/t=10$. Although $\Delta_{\rm SC}$ does not vary smoothly with system size, a linear extrapolation is plotted, providing rough approximations for the TL values of $\Delta_{\rm SC}$. The tensor network factor is not used in these calculations due to computational cost. Although the TL value of $\Delta_{\rm SC}$ is not precisely determined, it is clearly less than $\Delta_{\rm SC}$ in uniform states in the region $0.125\le\delta\le 0.1875$ (see Fig. \ref{f:uniform_sc_extrap}).  }
\label{f:stripe_sc_extrap}
\end{figure}

We obtain some evidence for nonzero SC in stripe states, however it is not conclusive. In Fig.~\ref{f:stripe_sc_extrap}, we plot $\Delta_{\rm SC}$ measured along a hole rich stripe as a function of system size for a $\lambda=6$, and $\lambda=7$ stripe, which was calculated using the mVMC wave function without the tensor-network factor. The stripe periods and dopings selected exibited large $\Delta_{\rm SC}$ in finite-size calculations. As can be seen $\Delta_{\rm SC}$ remains large for large systems and extrapolates to a nonzero value (albeit with large error bars). While these calculations suggest superconductivity may be present in stripe states, it is substantially less robust than in the charge uniform state. 

\section{Benchmark calculations}
\label{s:benchmarks}
\begin{figure}[t]
\includegraphics[width=0.4\textwidth]{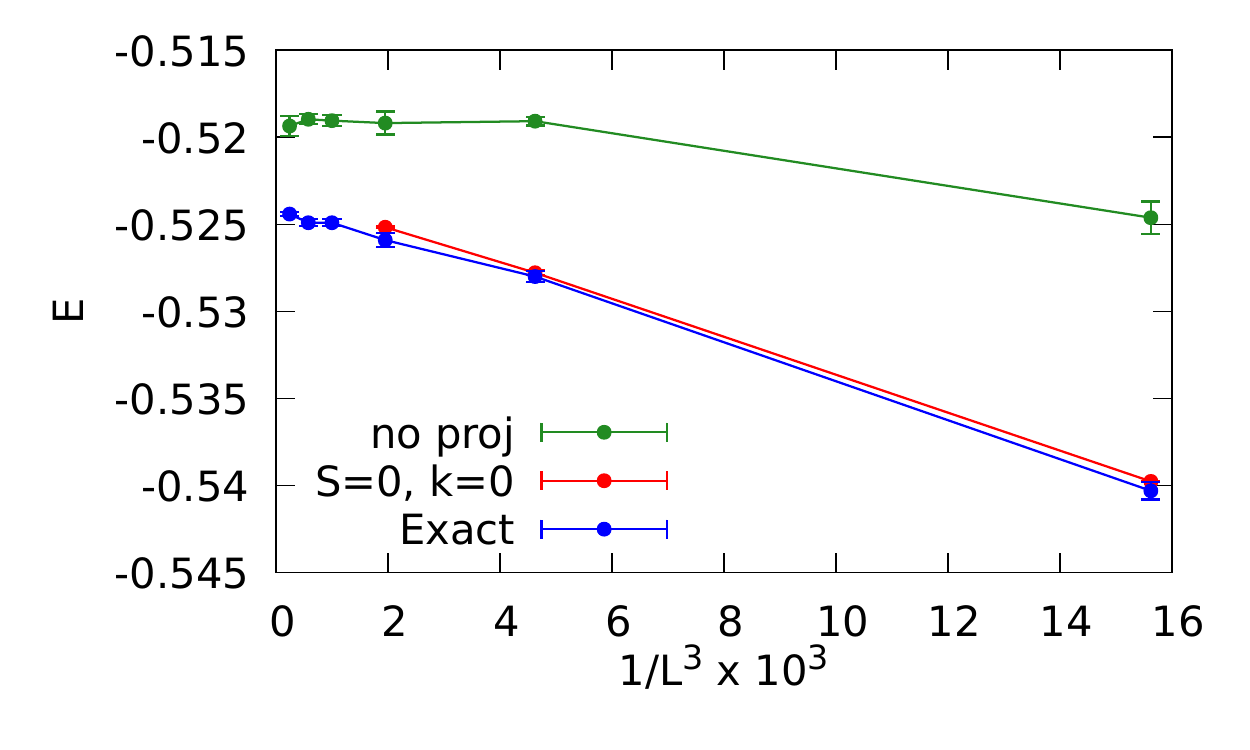}
    \caption{Comparison of extrapolated energies 
for $U/t=8$ at half-filling 
with numerically 
    exact QMC energies for various system sizes in units of $t$. The green line is without quantum number projections applied. The red line is obtained with $\mathcal{L}^K$ and $\mathcal{L}^S$ applied to $\psi_{\rm pair}(x)$ and $\mathcal{L}^K$  to $\mathcal{M}(x)$.  $2\times2$ unit cell is used in the calculations without quantum number projections, while the full unit cell is used when they are applied.}
\label{f:benchmark}
\end{figure}

We have performed a number of benchmark calculations to evaluate the accuracy of the method. At half-filling we compare our results to quantum Monte Carlo (QMC) data, which can be regarded as exact within the statistical error \cite{qin_benchmark_2016}. A comparison of our results with those obtained with QMC at $U/t=8$ are shown in Fig. \ref{f:benchmark}.  We have performed the calculations both with and without quantum number projections applied to the variational wave function. As seen in the benchmarks in Fig. \ref{f:benchmark} there is a discrepancy with the exact energies which decreases with system size to around $0.005t$ for a $16\times 16$ system when quantum number projections are not applied to the wave function. However, when $\mathcal{L}^K$ and $\mathcal{L}^S$ are applied to $\psi_{\rm pair}(x)$ and $\mathcal{L}^{C_4}$ to $\mathcal{M}(x)$, the discrepancy decreases to less than $0.001t$. Due to the large numerical cost, we did not apply quantum number projections to obtain the results in the main text. While the quantum number projections improve the energy, we show in Appendix \ref{s:projections} that they appear to have a relatively small effect on other physical quantities. 

Away from half-filling, exact results are not available. However, the 1/8 doping point at $U/t=8$ was recently studied using the density matrix renormalization group (DMRG), infinite projected entangled pair states (iPEPS), auxilliary field quantum Monte Carlo (AFQMC), and density matrix embedding theory (DMET) \cite{zheng_stripe_2017}. The different methods provided evidence of a stripe ground state with a near degeneracy of stripes with periods from 5-8.  We remark that, due to the extrapolations and approximations used, none of the results from these methods can be regarded as variational upper bounds to the true ground-state energy.  Therefore, a lower energy does not imply a more accurate method. However, the energies obtained using the different methods are close and varied within the range of  $0.01t$, indicating that the ground-state energy is likely around $-0.76t$ or $-0.77t$ with the uncertainty of $0.01t$. 

\begin{table}[!t]
\begin{tabular}{ | c | c | c | }
    \hline
    System size & Energy per site  & Energy per site  \\
    (unit cell)& TN+Lanczos & Var. Extrap. \\
    \hline
    \hline
    \multicolumn{3}{|c|}{No quantum number projection}\\
    \hline
    $14\times8 $ ($7\times2$)  &-0.7456(2)&-0.7539(4)\\
    $14\times16$ ($7\times2$) &-0.7446(2)&-0.7547(1)\\
    $28\times8$  ($7\times2$)&-0.7446(2)&-0.755(1)\\
    \hline
    \multicolumn{3}{|c|}{Quantum number projection}\\
    \hline
    $14\times8 $ ($7\times2$) &-0.7477(1)&-0.7562(2)\\
    $14\times16$ ($7\times2$) & -0.7449(2) &-0.7560(1)\\
    $14\times8 $ ($14\times8$)&-0.7508(1)&-0.7578(2)\\
    \hline
    \end{tabular}

    \caption{Benchmark calculations of a period 7 stripe at $\delta=0.125$. Lowest variational energy (using tensor network correlation factor and first-step Lanczos) and energy obtained by variance extrapolation with and without quantum number projections applied are shown per site in units of $t$. When quantum number projections are applied, $\mathcal{L}^K$ and $\mathcal{L}^S$ are applied to $\psi_{\rm pair}(x)$ and $\mathcal{L}^K$ is applied to $\mathcal{M}(x)$. Different unit cell sizes are indicated in parentheses. }
\label{t:bench_stripe}
 \begin{tabular}{| c | c |}
    \hline
    Method & Energy per site\\
    \hline
    \hline
      DMRG ($\lambda=7$)&-0.762(1)\\
      DMRG ($\lambda=8$)&-0.762(1)\\
      iPEPS ($\lambda=7$)&-0.763(2) \\
      iPEPS ($\lambda=8$)&-0.767(2) \\
      DMET ($\lambda=7$)&-0.7704(3)\\
      DMET ($\lambda=8$)&-0.7706(1)\\
      AFQMC ($\lambda=8$)&-0.7656(4)\\
      AFQMC ($\lambda=8$, $16\times 8$ PBC)&-0.7668(2)\\
      AFQMC ($\lambda=7$, $14\times 8$ PBC)&-0.7653(2)\\
    \hline
    \end{tabular}
    \caption{Energies per site, in units of $t$, obtained using four methods in Ref.~\onlinecite{zheng_stripe_2017} for $U/t=8$ at 1/8 doping for $\lambda=8$ and $\lambda=7$. Results represent approximations to TL values, except for the final two AFQMC entries, which correspond to finite-size systems with period boundary conditions. Note that these results are not variational, so lower does not necessarily imply more accurate.}
    \label{t:othermethods}
    \vspace{1em}

\end{table}

To compare with the above methods, we have calculated the energy of the $\lambda=7$ stripe at $U/t=8$ with $\delta=0.125$. The results of these calculations are included in tabular form in Table \ref{t:bench_stripe}. We have included results with various lattice and unit cell sizes, with and without quantum number projections. 

We observe a non-negligible improvement in the extrapolated energy when quantum number projections are applied to the variational wave function. This improvement is around $0.0015t$ for the $14\times 16$ system. Finite size effects are very small: after quantum number projections are applied, the extrapolated energies of $14\times8$ and $14\times 16$ systems are within error bars. Using a full lattice for the unit cell, rather than a $7\times 2 $ unit cell results in a further slight improvement of $0.002t$ to the energy when quantum number projections are applied. 

Comparing the energy with the results of Ref.~\onlinecite{zheng_stripe_2017} (shown in Table \ref{t:othermethods}), we see that the lowest energy obtained with our method is close but slightly higher than those obtained with other methods (about $0.003t$ above the error bars of DMRG and iPEPS). However, by considering the variation in the energy estimate among different methods and considering the non-variational aspects of several methods, our energy estimate is within the  uncertainty range, $0.01t$ in the previous study. In addition, the severe competition of the stripe states with the period of 5 to 8 lattice spacing is also consistent. 
Direct comparison of finite-size systems for benchmarking is difficult due to, for instance, the different boundary conditions required by different methods and other possible biases (e.g., the constrained path bias for AFQMC). We have, nevertheless, included finite-size results of AFQMC for comparison in Table \ref{t:othermethods}.
Additional benchmarks of the method at different $U/t$ can be found in Ref.~\onlinecite{zhao_variational_2017-1}.

\section{Quantum number projections}
\label{s:projections}

\begin{figure}[t]
\includegraphics[width=0.5\textwidth]{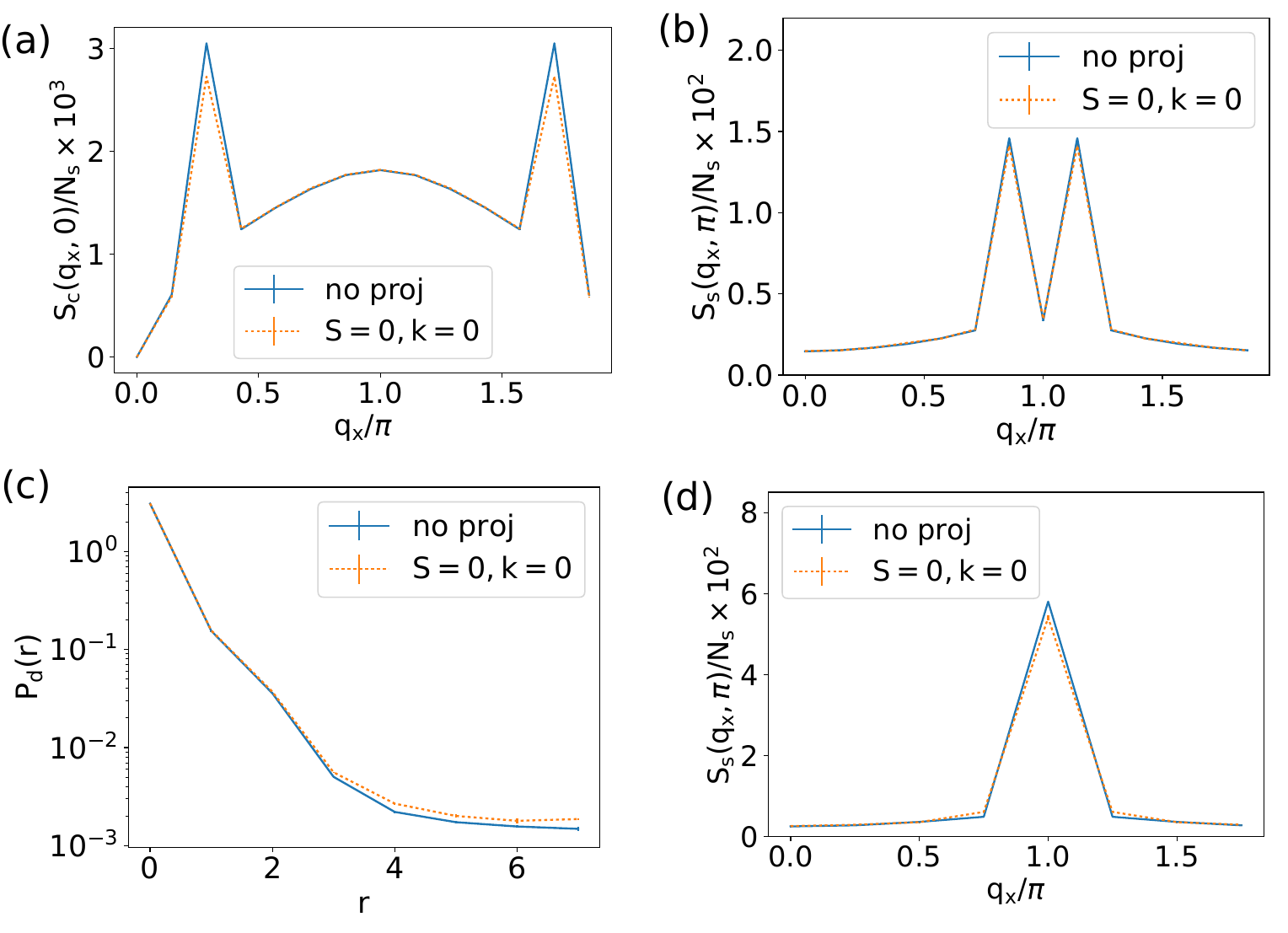}
\caption{Comparisons of different physical quantities with and without quantum number projections $\mathcal{L^K}$ and $\mathcal{L^S}$ applied to $\psi_{\rm pair}(x)$, which are denoted by ``S=0, k=0" and ``no proj" in the graph legends, respectively. Structure factor for (a) charge, (b) spin  and (c) superconducting correlations in the $x$ direction for a period 7 stripe at $U/t=8$ at $\delta=0.125$ on a $14\times8$ system. (d) spin structure factor at half-filling on a $8\times 8$ system at $U/t=8$.}
\label{f:chargesf_qnum}
\end{figure}

For finite-size systems, applying quantum number projections which restore translational symmetry, $SU(2)$ symmetry and space group $C_4$ symmetry can improve the energy of the wave function. This improvement is computationally expensive, so we have not used it in most of our calculations. While we have observed an improvement in energy when these projections are applied (see benchmarks in Appendix \ref{s:benchmarks}), the quantum number projections result in very little change in other physical quantities. The biggest improvement in energy was obtained by applying $\mathcal{L}^K$ and $\mathcal{L}^S$ to $\psi_{\rm pair}(x)$, which we test here. We have plotted charge structure factor, spin structure factor and superconducting correlations in stripe and uniform states with and without quantum number projections in Fig. \ref{f:chargesf_qnum}. As can be seen, the values of the physical quantities change little when quantum number projections are applied to the variational wave function.

\bibliographystyle{apsrev4-1}
%

\end{document}